# Document Classification Using Expectation Maximization with Semi Supervised Learning

Bhawna Nigam[1], Poorvi Ahirwal[2], Sonal Salve[3], Swati Vamney[4]

Department of Information Technology, IET, DAVV

[1]`bhawnanigam@gmail.com`  [2]`poorvi.ahirwal@yahoo.in`  [3]`salve_sonal@yahoo.com`
[4]`svamney@gmail.com`

*Abstract*

*As the amount of online document increases, the demand for document classification to aid the analysis and management of document is increasing. Text is cheap, but information, in the form of knowing what classes a document belongs to, is expensive. The main purpose of this paper is to explain the expectation maximization technique of data mining to classify the document and to learn how to improve the accuracy while using semi-supervised approach. Expectation maximization algorithm is applied with both supervised and semi-supervised approach. It is found that semi-supervised approach is more accurate and effective. The main advantage of semi supervised approach is "DYNAMICALLY GENERATION OF NEW CLASS". The algorithm first trains a classifier using the labeled document and probabilistically classifies the unlabeled documents. The car dataset for the evaluation purpose is collected from UCI repository dataset in which some changes have been done from our side.*

## 1.Introduction

Data mining [2][3] is the extraction of useful knowledge from large amount of data. Data mining tools can provide solution to the business problems that were to too time consuming when done manually. Classification[11] is one of the important aspect which comes under data mining and is a predictive modeling technique.

It is used to group the data or documents in some pre-defined classes i.e. to classify them according to attribute matching approach. Most companies have huge amount of data available which has to be refined, to apply classification on them. Data mining techniques can be implemented rapidly on existing software and hardware platforms to enhance the value of existing information resources, and can be integrated with new products and systems as they are brought on-line. Classification techniques [3] are used in various real world problems with respect to application domain as well as for various research purposes relevance to today's business environment as well as a basic description of how data warehouse architectures can evolve to deliver the value of data mining to end users. Data mining derives its name from the similarities between searching for valuable business information in a large database — for example, finding linked products in gigabytes of store scanner data — and mining a mountain for a vein of valuable ore. Both processes require either sifting through an immense amount of material, or intelligently probing it to find exactly where the value resides.

In general it is observed that nothing can be gained from the unlabeled data. But the fact is that they provide the join probability distribution of words.

This white paper represents EXPECTATION MAXIMIZATION ALGORITHM[6] with semi supervised approach that takes advantage of both labeled and unlabeled data. EM is a class of iterative algorithm for maximum likelihood and maximum posterior estimation in problem with unlabeled data.

## 2. Techniques Used for Classification

1. Artificial neural network.
2. Decision tree.
3. Expectation maximization algorithm.
4. Naïve baye's algorithm.

### 2.1. Description of document classification using Expectation algorithm

**2.1.1. Overview of Expectation Maximization**.

Create an initial model, $\theta_0$. Arbitrarily, randomly, or with a small set of training examples. Use the model $\theta'$ to obtain another model $\theta$ such that

$\Sigma_i \log P_\theta(y_i) > \Sigma_i \log P_{\theta'}(y_i)$

Repeat the above step until reaching a local maximum. Guaranteed to find a better model after each iteration.

**2.1.2. Steps for preprocessing and classifying document[7] can be summarized as follows:**

- Remove periods, commas, punctuation, stop words. Collect words that have occurrence frequency more than once in the document. We called this collection of words as vocabulary.
- View the frequent words as word sets by matching the words which are in the vocabulary as well as training set documents.
- Search for matching word set(s) or its subset(containing items more than one) in the list of word sets collected from training data with that of subset(s) (containing items more than one) of frequent word set of new document.
- Collect the corresponding probability values of matched word set(s) for each target class.
- Calculate the probability
- Apply z score algorithm to calculate the range in which the attributes must be lying.
- Calculate the probability class by applying expectation maximization algorithm.
- Categorize the document in the class having maximum probability.

**2.1.3. BASIC EM**

The Expectation-Maximization algorithm is used in maximum likelihood estimation where the problem involves two sets of random variables of which one, X, is observable and other, Z, is hidden. In simpler words algorithm works in following two steps:

**E-step**

Estimates the expectation of the missing value i.e. unlabeled class information. This step corresponds to performing classification of each unlabeled document. Probability distribution is calculated using current parameter.

**M-step**

Maximizes the likelihood of the model parameter using the previously computed expectation of the missing values as if were the true ones.

Step 1-

Given:

  X - Labeled data

  Z - Missing values

  θ - Unknown parameter

$L(\theta; X, Z) = p(X, Z | \theta)$ - likelihood function (probability).

$L(\theta | X) \in \{\alpha\, p(X | \theta) : \alpha > 0\}$ , $p(X | \theta) * p(Z | \theta)$

Step 2:

With the given variables X, Z and θ, the maximum likelihood estimation of the unknown parameters is calculated by the marginal likelihood of the observed data. The value obtained is not tractable.

Finding maximum likelihood

**$L(\theta; X) = p(X | \theta) = \sum_{z} p(X, Z | \theta)$**

Step 3- caculate expected value of log likelihood function.

**$Q(\theta | \theta^{(t)}) = E_{z|x, \theta^{(t)}} [\log L(\theta; X, Z)]$**

 Step 4- find the parameters that maximizes this quantity

**$\theta^{(t+1)} = \arg\max Q(\theta | \theta^{(t)})$**

The observed data points X, may be discrete which are finite or taken from countably infinite set or it may be continous which are taken from uncountably infinite set.

The EM algorithm can be applied to other data models when one of the parameters Z or θ is known. This shows the iterative nature of algorithm that continues to predict the values until converges to a particular and specific value.

## 3. Applying Expectation Maximization algorithm with Semi Supervised Approach

We applied semi supervised approach for increasing the accuracy and to reduce the human efforts for classifying the document. Until now EM is applied with supervised approach, but the disadvantage of this is that the whole data must be labeled, and no dynamical generation of new classes would be there, which is too time consuming and leads to reduction in speed of classification.

In semi supervised approach we have both labeled and unlabeled data. From labeled data we give training to data set and then the unlabeled are classified. Submitted document is categorized in some pre defined classes and incase the document fails to lie in those defined classes then dynamically a new class is generated and document is categorized in that class and that new class is updated in database.

## 4. Implementation:

We are taking "car evaluation" dataset from **UCI repository**. We are considering 5 attributes buying, maintenance, , price, mileage and safety and three pre-defined classes unacceptable, good and very good. There are 1500 rows in the dataset. Half of the data is considered to be the training data and remaining the testing. These 5 attributes are used to classify the document submitted by user. Before the unlabeled document is read, two check are made. First the format of the document is checked, if its .txt, then only its is processed further otherwise declared as invalid. If the document passes the first check it must also lie only in the car domain to be classified. If it is within domain then further steps of classification are done else process is ended there only. The document to be classified is read and the classifier searches these five attributes in that document. After mining these attributes from document the classifiers calculates the probability of that document to belong to our pre-defined classes. We make a set of probabilities which is the initial set. According to the probability calculated the classifier expects the class of document i.e. E –step is done. Now use that set to create a subset of that provided the probability of subset is maximum then the previous set. This procedure is done iteratively until and unless we reaches at the maximum probability of falling any document in pre-defined classes. Classification is predicated on the idea that electronic documents can be classified based on the probability that certain keywords will correctly identify a piece of text document to its annotated category. Document is classified in the correct class and if the probability does not match with any of these classes, a new class is dynamically generated and updated in the dataset.

## 5. Comparison of Expectation Maximization with Supervised and Semi Supervised Approach

Disadvantage of supervised approach with expectation maximization algorithm is that, first of all the data should be labeled hence increases the efforts and time consuming and the unlabeled data is useless here.

In case the document fails to lie in pre defined classes the approach is not able to classify that document hence leading to decrease in efficiency of classifier.

In semi supervised approach unlabeled data is not useless, it has been used to train the classifier and in case the document fails to lie in pre defined classes it leads to dynamically generation of new class to categorize that document and the database is updated automatically.

International Journal on Soft Computing ( IJSC ) Vol.2, No.4, November 2011## 6. Performance Analysis

For measuring the performance, the following definitions of precision, recall, accuracy and F1-measure are used to find the effectiveness of document classifier. Accuracy of classifier is the percentage of documents correctly classified by classifier

$$Accuracy = \frac{Number\ of\ correct\ prediction}{Total\ number\ of\ prediction}$$

Recall is determined by number of documents retrieved that are relevant with respect to total number of documents that are relevant. Thus,

Recall=TP/(TP+FN)

Precision is determined by number of documents retrieved that are relevant with respect to total number of documents that are retrieved. Thus,

Precision=TP/(TP+FP)

Thus,

F1-measure=(2*recall*precision)/(recall+precision)
Where,

TP (True Positive): The number of documents correctly classified to that class.

TN (True Negative): The number of documents correctly rejected from that class.

FP (False Positive): The number of documents incorrectly rejected from that class.

FN (False Negative): The number of documents incorrectly classified to that class.

## 8. Experimental Results

The work of classifying a new document depends mainly on the matching of values of attributes with those of the values of predefined classes. We studied and implemented both the approaches discussed above. For performance evaluation, we used the Accuracy, Precision and Recall metrics that were presented in the previous section. In the following two figures both the techniques are compared with each other. The comparison is done on the basis accuracy and F1-Measure calculated for both the techniques. Accuracy and F1-Measure are compared with respect to number of documents given to classifier.

As observed in the experiments, semi supervised approach of classification is ahead of supervised method. Also, the dynamic generation of new class and updation in the database increases the accuracy and precision of the result.

The results obtained on performing the experiments are:

.

41

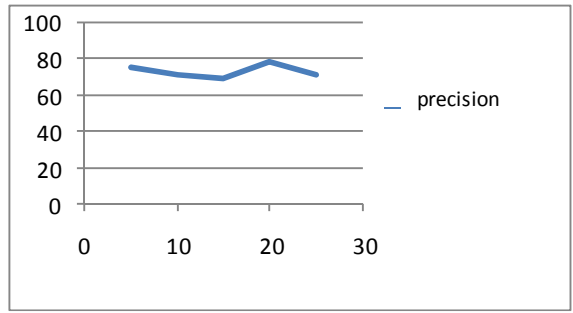

Fig1. Experimental results

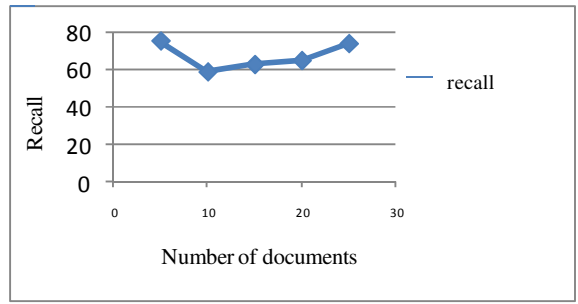

Fig2. Experimental results

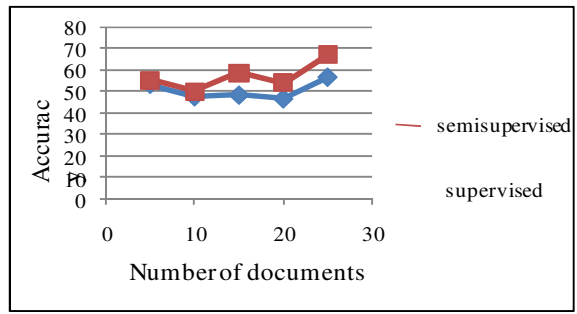

Fig3. Experimental results

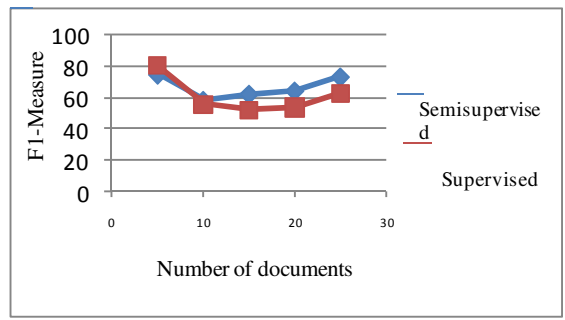

Fig4. Experimental results

## 9. Conclusion and Future Work

This paper has presented an algorithm that address the question of when and how unlabeled data may be used to supplement scarce labeled data, especially when learning to classify text document s. This is an important question in text learning because of the high cost of hand labeling data and because of the availability of large volumes of unlabeled data. We have presented an algorithm that takes advantage of it and experimental results that show significant improvements by using unlabeled documents for training classifier in three real-world document classification tasks.

When our assumption of data generation are correct, basic EM can effectively in corporate information from unlabeled data. However, the full complexity of real-world text data cannot be completely captured by known statistical models. It is interesting then, to consider the performance of a classifier based on generative models that make incorrect assumptions about the data. In such cases, when the data is inconsistent with the assumptions of the model, our method for adjusting the relative contribution of the unlabeled data prevents the unlabeled data for degrading classification.